%% file: paper.tex
\documentclass{egpubl-arxiv} 
\usepackage{eg2022}
\ConferencePaper

\usepackage[T1]{fontenc}
\usepackage{dfadobe}  
\usepackage{flushend}

\electronicVersion

\usepackage[pdftex]{graphicx} \pdfcompresslevel=9

\usepackage{xcolor}
\definecolor{someColor}{HTML}{FF7744}

\usepackage[caption=false,font=footnotesize,labelfont=sf,textfont=sf]{subfig}
\usepackage{graphicx}
\usepackage{enumitem}
\usepackage{booktabs}
\usepackage{hyperref}
\usepackage{xspace}
\usepackage{comment}
\usepackage{tabularx}

\input{commands}

\usepackage[nolist]{acronym}
\begin{acronym}
\acro{NN}{Neural Network}
\acro{CNN}{Convolutional Neural Network}
\acro{MSE}{Mean-square Error}
\acro{GAN}{Generative Adversarial Network}
\acro{LSGAN}{Least-squares GAN}
\end{acronym}

\newcommand{\network}[1]{\mathcal{#1}}
\newcommand{\image}[1]{{#1}}
\newcommand{\old}[1]{{#1}_\mathrm{o}}
\newcommand{\new}[1]{{#1}_\mathrm{n}}

\mymath{\lightDirection}{\omega}
\mymath{\colorImage}{\image{C}}
\mymath{\depthImage}{\image{Z}}
\mymath{\normalImage}{\image{N}}
\mymath{\shadowImage}{\image{S}}

\mymath{\relightingNet}{\network{L}}
\mymath{\cshadowNet}{\network{S}}
\mymath{\sshadowNet}{\network{D}}
\mymath{\advNet}{\network{A}}
\mymath{\refineNetwork}{\network{R}}
\mymath{\depthNet}{\network{Z}}
\mymath{\loss}{f}
\mymath{\lossWeight}{\lambda}
\mymath{\tunableParameters}{\theta}
\mymath{\imageMetric}{\Delta}

\newcommand{\method}[1]{\textsc{#1}}

\newcommand{\mypara}[1]{\noindent\textbf{#1}\ }

\title{OutCast: Outdoor Single-image Relighting with Cast Shadows}

\author[D. Griffiths, T. Ritschel, J. Philip]{
 David Griffiths$^{1,2}$\qquad 	
 Tobias Ritschel$^2$\qquad
 Julien Philip$^1$
 \\
 $^1$Adobe Research\qquad		
 $^2$University College London
 }

\teaser{
  \centering
  \includegraphics[width=\textwidth]{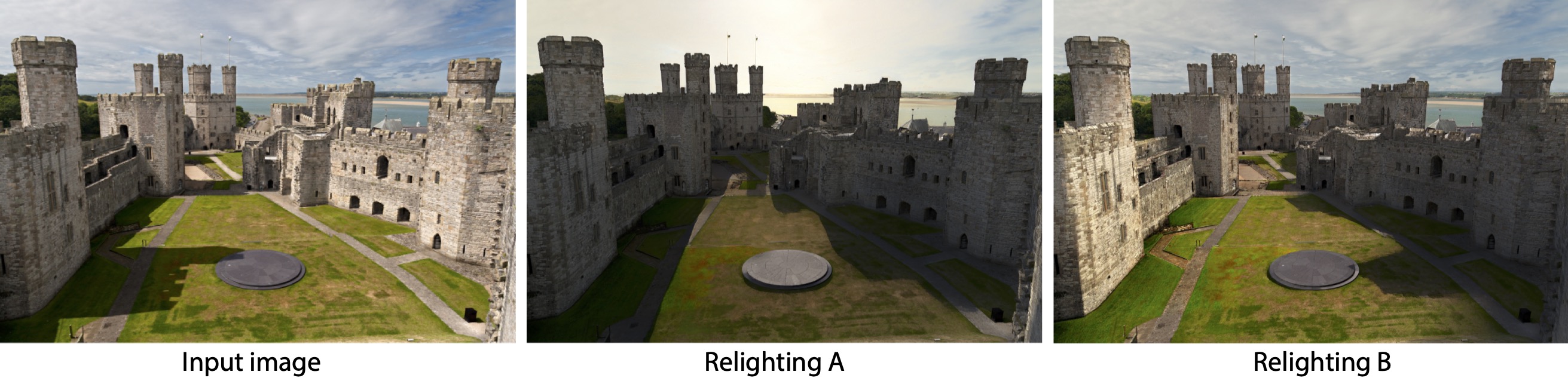}%
  \caption{Given only a single RGB image \emph{(left)} in one lighting, our method generates images of that scene in new lighting \emph{(middle / right)}.}%
  \label{fig:teaser}
}%

\begin{document}
\maketitle

\begin{abstract}
We propose a relighting method for outdoor images. 
Our method mainly focuses on predicting cast shadows in arbitrary novel lighting directions from a single image while also accounting for shading and global effects such the sun light color and clouds.
Previous solutions for this problem rely on reconstructing occluder geometry, \eg using multi-view stereo, which requires many images of the scene.
Instead, in this work we make use of a noisy off-the-shelf single-image depth map estimation as a source of geometry.
Whilst this can be a good guide for some lighting effects, the resulting depth map quality is insufficient for directly ray-tracing the shadows.
Addressing this, we propose a learned image space ray-marching layer that converts the approximate depth map into a deep 3D representation that is fused into occlusion queries using a learned traversal. 
Our proposed method achieves, for the first time, state-of-the-art relighting results, with only a single image as input.
For supplementary material visit our project page at: \href{https://dgriffiths.uk/outcast}{$\mathtt{https://dgriffiths.uk/outcast}$}.

\ccsdesc[300]{Rendering ~ Relighting}
\printccsdesc   

\end{abstract}

\mysection{Introduction}{Introduction}
Capturing stunning photographs requires a subtle equilibrium between the subject, the composition and the lighting of a scene.
While a user can decide what subject to capture and control the properties of the sensor, obtaining the right lighting is much more challenging and often either requires patience and dedication, or is simply out of the user's control.
For outdoor pictures, where the sun and sky lighting is dominant, previous methods have proposed to relight an image by removing and re-synthesizing shadows \cite{yu1999inverse,tchou2004unlighting,duchene2015multiview,philip2019multi}, this problem is well-defined and known to be notably challenging in computer graphics literature.
One of the key hurdles is that occluding geometry casting shadows can be arbitrarily far away from the point receiving the shadow, thus requiring a fine understanding of long-range interactions between objects is necessary.
The most dramatic shadow shots \eg a sunset illuminating architecture, are notoriously difficult in this respect as the shading of a point can depend on arbitrarily far geometry. In addressing this issue, we propose a method that takes a single RGB image as input and enables a user to change its illumination, including dominant cast shadows.

When accurate 3D geometry of a scene is available, cast shadows can be computed precisely using ray-tracing or shadow mapping.
Alternatively, given multiple photos from varying view-points of the scene, 3D proxy geometry can be estimated. Such geometry, even if approximate, has been proven sufficient to produce faithful shadows and global illumination effects \cite{philip2019multi,philipp2021free}.
A challenge in these approaches is to adapt the light transport computation to become robust to the approximate 3D geometry, \eg by using a \ac{NN} to combine shadows with the actual image (\refFig{Contribution}, gray arrows).
In this paper we go to the extreme, and show for the first time, how to cast shadows from very approximate and incomplete geometry (a depth map), extracted from a single RGB image alone (\refFig{Contribution}, black arrows).
To do so we demonstrate how to leverage off-the-shelf NN-based depth maps \cite{eigen2015predicting,ranftl2020midas,Wei2021CVPR} and the limited geometric information they provide to compute plausible cast shadows and shading for an arbitrary sun position.

We achieve this by combining classic screen space shadowing \cite{ritschel2009approximating} with a learned component to produce both attached and cast shadows as well as more accurate shading.
Our learned component is convolutional and able to attend screen space information relevant to casting a shadow, while conventional image-to-image translation \cite{zhu2017unpaired} from depth to shading \cite{nalbach2017deepshading} is unable to deal with such long-range interactions.

Our main contribution in this work is thus a new hybrid component mixing image space graphics and learnt priors. This component provides a way to compute precise long-range interaction, such as cast shadows, from depth maps alone. This allows plausible relighting in challenging outdoor environments from a single picture.

\myfigure{Contribution}{Our method \emph{(black arrows)} extracts a depth map from a single view to compute cast shadows while previous work \emph{(grey arrows)} relied on a global multi view-generated 3D proxy.}

\mysection{Previous Work}{PreviousWork}

Our work builds on research in both the fields of image lighting estimation/relighting and deep learning-based methods for image-to-image tasks. 
For a general survey on (deep) lighting estimation and relighting we refer the reader to \cite{einabadi2021deep}.

\mysubsection{Lighting and Shading Estimation}{}
A key aspect to changing the lighting of an image is understanding its original lighting conditions. 
For example, the source image shading can inform a relighting algorithm which shadows need to be removed in a target image, or give important cues regarding the light source. 
Early works on image-based lighting \cite{debevec2002image} demonstrated the ability to use captured lighting from images covering the hemisphere to render synthetic objects under novel lighting conditions, a technique which was further extended to high dynamic range images \cite{stumpfel2004direct}. 
In contrast, many recent works attempt to directly estimate a parametric lighting representation from a single RGB image. 
Other work \cite{lalonde2009estimating} exploits cues extracted from varying portions of the image (\eg sky, vertical surfaces and ground), as well as shadows, shading and approximate geometry to estimate the sun position, which can in turn be used to generate a synthetic sun dial.

Recent works \cite{hold2017deep,holdgeoffroy-cvpr-19,zhang-cvpr-19} leverage advances in deep learning to estimate outdoor sky parameters.
CNN-based architectures are shown to be effective in estimating high dynamic range or parametric outdoor illumination from low dynamic range image inputs  \cite{legendre2019deeplight}. 
The resulting environment maps can effectively be used to directly render new synthetic objects into the original image, as long as complex interactions between shadows are avoided.
The approaches mentioned all assume an outdoor sky model with a single source of light for illumination. 
Other work \cite{Gardner_2019_ICCV} proposes a method to estimate lighting of indoor scenes for a single image with multiple light sources of varying properties. 
The authors achieve this by defining a parametric lighting representation describing area lights distributed in 3D for each pixel making the method computationally cumbersome, preventing the computation of hard shadows due to the low resolution of the representation. 
 Similar work \cite{li2020inverse} proposed an indoor network which can estimate scene shape, spatially varying lighting (driven by a spherical Gaussian lighting representation) and non-Lambertian surface reflectance.

These major steps forward in lighting estimation, unfortunately, only formulate part of the solution to the relighting problem.
For instance,  Li et al., \cite{li2020inverse} estimate intrinsic parameters such as normals and albedo, allow object insertion and some level of scene editing but not relighting. 
Even with this rich information, altering the entire image illumination is not  straightforward, as one also needs to define a coherent lighting, shading and shadows for the novel illumination.

Another essential component of the relighting problem is the automatic removal of source shadows. Even with access to accurate shadow masks, this is a challenging task. 
Such algorithms must adjust the shadow pixel intensities, whilst also inferring semantic understanding of the scene to handle fine shading details. 
DeshadowNet \cite{qu2017deshadow} is a multi-branch CNN which learns both local and global features of the input image.
Wang et al., \cite{wang2018stacked} propose a novel approach where separate shadow detection and removal networks can mutually benefit from each other by introducing adversarial losses.
In this work we jointly perform shadow removal, re-casting and relighting in a single unified network. 
Whilst this demands considerably more from the network, we demonstrate empirically these tasks can be learned together.

Intrinsic decomopsition of images can also be a useful technique for extracting shadows from RGB images \cite{bonneel2017intrinsic}.
Such methods can identify (and subsequently remove \cite{qu2017deshadownet}) source shadows and shading in the scenes and can now run at an interactive frame rate \cite{meka2021real}.
Many early works adopted this approach \cite{isaza2012evaluation} to enable some form of lighting editing.
Unfortunately, it is not obvious how these methods can be extended to produce the inverse, shadow generation, especially with long range interaction.

\mysubsection{Relighting}{}

Prior work on image relighting typically relies on scene geometry, light and reflectance models to accurately relight using inverse rendering \cite{yu1999inverse,loscos1999interactive,marschner1997inverse}.
Having access to the full scene representation (geometry, materials, lighting) allows traditional rendering and shading methods to be used and gives promising results, however, requires a highly complex capture process.
The capture process can be simplified through techniques such as semiautomatic vision-based geometry reconstruction \cite{loscos2000interactive}, or by computing scene parameters through viewing the same scene under varying lighting conditions \cite{eisemann2004flash,loscos1999interactive}, however, a high level of technical competency is still required.
Furthermore, lighting and material information requirements can be relaxed using Inverse Path Tracing \cite{azinovic2019inverse}, however, a full and accurate scene geometry is still essential for high quality rendering.
If very high quality results are required (\eg for film production), sophisticated capture sets ups such as the Light Stage \cite{debevec2000acquiring,wenger2005performance,meka2019deep,relightables,sun2020light} can be employed.
Whilst such methods give impressive results, they require expensive rigs and trained professionals to operate them.

Advancements in deep learning have resulted in framing full scene representation as an optimization problem \cite{sitzmann2019siren,mildenhall2020nerf}.
An implicit representation of the scene is jointly optimized for geometry but can also be optimized for surface characteristics such as albedo and normals, enabling re-rendering under new lighting conditions using traditional rendering pipelines \cite{DBLP:journals/corr/abs-2008-03824,nerfactor,boss2021nerd,DBLP:journals/corr/abs-2007-09892,nerv2021}.
More recent work such as PixelNeRF \cite{yu2021pixelnerf} show that with appropriate conditioning reasonable results can be obtained from single-image inputs.
Unlike our method, these works are restricted to objects, rather than entire scenes and require multiple views as input or require strong preconditioning and assumptions of the scene.

For scene scale relighting, deep learning has enabled significant advances in the performance of multi-view relighting systems \cite{philip2019multi,philipp2021free}.
Typical approaches employ a \ac{NN} to map from one of the input images and a set of approximate guide maps (depth, normals, shadow images) to a novel illumination condition. 
2D image \ac{NN}s are used to jointly remove the old shadow and shading and change it to the new lighting and shadow without attempting to recover an explicit shadow-free image \cite{sanin2012shadow} or an albedo map \cite{land1971lightness}, requiring only geometry to guide the relighting. 
However, the quality of the results still comes at the cost of the capture process.
To allow for a proxy geometry, photogrammetric pipelines can be utilized, requiring tens or hundreds of images of the scene from varying view-points. 
In many practical applications, multi-image data acquisition is not possible.
Whilst impressive results have been achieved using only image sets with as little as five images \cite{xu2018deep, ren2015image}, these approaches are typically constrained to objects or simple scenes.

It is specifically this constraint we aim to relax in our work.
We build on the work of Philip et al,. \cite{philip2019multi}, however, only require a single-image input. 
Despite this, we achieve a visually comparable performance on challenging real-world test cases (\refFig{Comparison}).
Key to our method is the use of an off-the-shelf depth estimator \cite{eigen2015predicting,ranftl2020midas,Wei2021CVPR} as a source of approximate geometry and a novel 3D module to become robust to such inputs.

Single-image relighting is not a new problem. 
Many studies focus on relighting limited subjects such as human faces \cite{peers2007post,wang2009face,shu2017neural,iccvportrait,sun2019single,nestmeyer2020faceRelighting} and bodies \cite{kanamori2018relighting,Lagunas2021humanrelighting}. 
On these restricted classes deep priors are easier to build, allowing for very impressive results such as recently demonstrated in Total Relighting \cite{Pandey:2021}.
Regarding more general scenes, Wu et al., \cite{wu2017interactive} obtain realistic relighting results from single images, but at the cost of significant user interactions to annotate the scene and estimate the geometry. 
Ture et al., \cite{ture2021cgf} focus mostly on sky relighting. 
Similar to our method, monocular depth estimation is used, however, their approach can only handle shadows cast by clouds. Single-view relighting methods \cite{yuye2020,liu2020factorize} have also recently built on image-to-image translation methods \cite{zhu2017unpaired,wang2018pix2pixHD}. 
One of the key missing components of these methods is their ability to generate accurate and convincing cast shadows in the target image which is our main contribution. 
Liu et al., \cite{liu2020factorize} show some level of shadow casting, but these are overly smooth and soft and the method is restricted to cities.

Other works deal with shadows by assuming video / time-lapse input \cite{sunkavalli2007factored}, terrain data \cite{kopf2008deep} or object templates \cite{kholgade2014object,karsh2011rendering}. 
The outdoor scenes and the capture modality we target do not match such requirements.
Attempts at predicting cast shadows are also made using traditional 2D CNNs \cite{carlson2019shadow,zhang2019shadowgan,liu2020arshadowgan,Sheng_2021_CVPR}, allowing to cast approximate shadows of individual objects with limited quality especially when fine, long-range, interactions are needed, such as with hard cast shadows.
Still, we study 2D CNNs as a baseline for our method, demonstrating they are insufficient to solve our task.
This is because a typical U-net \cite{ronneberger2015u}, or even more advanced architectures \cite{wang2018pix2pixHD}, while being able to aggregate information at multiple scales, have no inductive bias to attend the information for long-range shadow interactions, which is in the direction of the light.

\mysubsection{Epipolar Geometry}{}
The shortcoming of 2D \acp{CNN} to explicitly collect relevant features across an image has been identified in prior work. 
Most notably epipolar Transformers \cite{he2020epipolar} create feature volumes by sampling along the Epipolar line of 2 images of the same scene with a known transformation. 
Explicitly sampling along a known ray boosted performance for 3D human joint localization. 
Similar conclusions have been drawn for depth regression \cite{prasad2018epipolar}, data-adaptive interest points \cite{yang2019learning} and keypoint detection \cite{jafarian2018monet}.
Shin et al., \cite{shin20193d} further adapt epipolar transformers for the task of 3D scene reconstruction for single-view RGB images.
Our 3D shadow network (\refFig{Marching}) takes inspiration from such networks.
However, instead of our sampling direction being determined by epipolar geometry, the direction is determined by the position of the light source. 
Furthermore, the above methods sample in feature space, whereas we directly sample the input RGBZ image.

\mysection{Background}{Background}

Our work is based on a method proposed by Philip et al., \cite{philip2019multi}, which relights an RGB image $\old\colorImage$, captured in an (unknown) original light condition characterized by the sun direction $\old\lightDirection$, to a novel light direction $\new\lightDirection$.
A particular strength of their method is the ability to render accurate cast shadows.
This is achieved through the use of \emph{shadow images} which are represented as gray-scale images $\old\shadowImage$ and $\new\shadowImage$ that hold the shadow information in the old and new light direction, respectively.
These intermediate maps are not used in a classic inverse rendering setting, but instead serve as guides to an image-to-image translation network to perform the final relighting.
However, constructing these shadow images relies on global 3D geometry, acquired with multi-view reconstruction \cite{ullman1979interpretation}, necessitating tens to hundreds of images.

In this work, we keep the overall structure of the system propsed by Philip et al., but relax the requirement of multi-view images as input.
Instead, we produce the shadow images from the original color image $\old\colorImage$ alone.
To enable this, we assume access to an approximate depth estimation process $\depthNet(\old\colorImage)$ with only scale-invariance, \eg an  off-the-shelf NN \cite{Wei2021CVPR}.
Alternatively, depth estimation could come from an active depth sensor \cite{zhang2012microsoft, arkitscenes}, which are becoming increasingly popular on smartphones.

Casting shadow from such approximate geometry, is more challenging than casting shadows from exact geometry via shadow mapping or ray-tracing.
This is largely due to high levels of occlusions, resulting in incomplete geometry.
A depth map only provides the geometry of the visible surface when computed from a single view-point of the scene.
A depth map does not provide information regarding what is behind an object thus direct shadow casting across its back will incorrectly report in no shadows (\refFig{DirectProblem}) or too much shadows (\refFig{Problems}, second column).
Our main contribution is a learnable module that takes as input samples obtained from a depth-based shadow casting approach, and outputs a shadow mask, robust to approximate and incomplete geometry.

\mycfigure{Architecture}{
The main network architecture of our approach.
Input is outlined in orange, output in green.
Pink lines indicate the cost functions we optimize for.
Blocks with learnable parameters are double-wedges, and a dotted line indicates siamese training / weight sharing.
The scissor symbol is where the gradients are detached from the network graph, to prevent further back propagation of the gradient.}

Formally, the system proposed by Philip et al., \cite{philip2019multi} is a \emph{relighting operator} $\bar\relightingNet(
\old\colorImage,
\old\shadowImage,
\new\shadowImage,
\old\lightDirection,
\new\lightDirection
)$.
Where $\old\colorImage$ is the image which we want to relight,
$\old\shadowImage$ and $\new\shadowImage$ are the respective old and new shadow images computed thanks to the mulit-image derived 3D proxy geometry, and where $\old\lightDirection$ and $\new\lightDirection$ represent the old and new sun directions respectively.
In this work we focus on removing the requirement for the multi-image derived 3D geometry proxy that allows the computation of old and new shadow images $\old\shadowImage$ and $\new\shadowImage$ as system inputs, resulting in a new operator $\relightingNet(
\old\colorImage,
\old\lightDirection,
\new\lightDirection
)$,
relying only on the color image $\old\colorImage$ and the input light direction $\old\lightDirection$.
Details on how we obtain $\old\lightDirection$ can be found in \refSec{Deshadow}.

\mysection{Image Relighting using Approximate Depth Maps}{OurApproach}

Our new relighting approach (\relightingNet) shares the high level architecture of the one proposed by Philip et al., \cite{philip2019multi} ($\bar\relightingNet$) using shadow images directly computed from the color image $\old\colorImage$ alone as input: \[
\new{\colorImage}=
\relightingNet(
\old{\colorImage},
\old\lightDirection,
\new\lightDirection)
\simeq
\bar\relightingNet(
\old\colorImage,
\cshadowNet(\old{\colorImage}, \old\lightDirection), 
\sshadowNet(\old{\colorImage}, \new\lightDirection),
\old\lightDirection,
\new\lightDirection)
),
\]
where \cshadowNet produces the new shadow image (\refSec{Reshadow}) and \sshadowNet produces the old shadow image (\refSec{Deshadow}) that are combined in the final relighting step (\refSec{Relighting}).
\refFig{Architecture} shows an overview of our approach.
\refFig{Buffers} shows an example of the intermediate buffer maps on test data. 

\mypara{Depth and normal estimation}
We make use of a depth estimation method based on MiDaS \cite{ranftl2020midas}, denoted as
$\depthImage=\depthNet(\colorImage)$. 
As the initial MiDaS implementation is trained to produce disparity maps that are not only scale-invariant, but also shift-invariant, an unknown non-linear distortion would be applied to depths values if we were to use it directly, leading to poor normals and distorted geometry \cite{Wei2021CVPR}.
We therefore use a version of MiDaS \cite{ranftl2020midas} trained without using the shift-invariant losses so it recovers the scene with an unknown scale but no additional shift.
This is achieved by training on data that is either synthetic, LiDAR sensed, or from stereo camera pairs with known calibration, meaning there is no shift ambiguity in the data.
We do not refine the weights of \depthNet during training.

We compute an approximate normal image $\normalImage$, by first converting the depth map to a 3D position map containing for each pixel its $x,y,z$ coordinates according to the camera frame of reference (operator $\rho$) and then taking the normalized cross product between the horizontal and vertical components, $u$ and $v$, of the gradient of the position map.
This gradient is computed using a Sobel filter.
\[
\normalImage=
\frac{\partial \rho(\depthImage)}{\partial u}
\times
\frac{\partial \rho(\depthImage)}{\partial v}
\]
This process does not need to be differentiable, it only allows to use an approximate depth \depthImage or normals \normalImage whenever working with a color image \colorImage.
All operations are performed in camera space.

\mypara{Loss}
Our relighting $\relightingNet_\tunableParameters$ is a \emph{tunable mapping}, a function with learnable parameters \tunableParameters which are trained end-to-end, minimizing a cost function of four terms: \begin{equation}
\loss=
\lossWeight_\cshadowNet
\loss_\cshadowNet+
\lossWeight_\sshadowNet
\loss_\sshadowNet+
\lossWeight_\relightingNet
\loss_\relightingNet+
\lossWeight_\advNet
\loss_\advNet
.
\end{equation}
where $\lambda$ scales each loss, respectively.
We will explain the new-shadow, old-shadow, relighting and adversarial loss terms 
$\loss_\cshadowNet$,
$\loss_\sshadowNet$,
$\loss_\relightingNet$
and
$\loss_\advNet$
in the following respective Sections \ref{sec:Reshadow}, \ref{sec:Deshadow} and \ref{sec:Relighting}.

\mysubsection{Producing a new shadow}{Reshadow}
\myfigure{Problems}{Illustration of the inaccuracy and the incompleteness problems when casting shadows from a depth map.
The top row shows a scene in ideal conditions with complete global geometry.
The bottom row shows shadows resulting from a high threshold direct shadow casting with approximate depth map geometry.}

\mypara{Problem analysis}
The main challenge we address in this paper is how to cast shadows using an imperfect depth map.
A depth map differs from the global proxy 3D geometry in two important aspects: it is more inaccurate and it is incomplete.
Therefore, directly applying shadow casting to obtain detached shadow-images will produce unsatisfactory results.

It is \emph{inaccurate}, because a NN-based depth estimator is never able to perfectly match the true depth.
It lacks details and often suffers from distortion.
For example, an even ground plane or a side wall of a building facing away from the light typically does not come out as a proper plane, rather it would appear bumpy and curved.
These effects are particularly strong for texture objects in a phenomenon known as ``texture copying''.
Such bumps, when used to compute attached shadows will lead to several false positives: normals computed from bumpy depth will make surfaces sporadically face away from the light, \ie darker.
When used during the computation of detached shadows, they cast small shadows known as ``shadow acne'' in the shadow mapping literature \cite{dou2014adaptive}, this phenomenon is illustrated in the first column of \refFig{Problems}.
Severity of these difficulties varies from method to method and from sensor to sensor but are clearly present in the state-of-the-art depth estimator we use in our experiments as well as in low-cost active depth sensors (\eg LiDAR sensors).

Secondly, even if an oracle NN would reconstruct depth values perfectly, these will remain \emph{incomplete}, because a 2D depth map can only store a single depth for every pixel.
If we see a frontal RGB image of a box, we cannot know how far it extends behind every pixel.
This information, however, is critical to cast a proper shadow.
A thin box will cast a thin shadow, a thick box will cast a thick shadow.
Hence, ideally we would not only need to know the depth of the surface, but also have a notion of thickness.
This is illustrated in \refFig{Problems} (Col. 2), where we visualize a shadow obtained by naively intersecting a surface created by a depth map if we assume all geometry extends indefinitely away from the camera. We show the shadow obtained by naively intersecting the surface created by the depth if we meshed it, that is, whenever a ray pass behind an observed surface it is considered in shadow. 
Other possibilities to compute shadows from depth maps using thresholding are shown in \refFig{Threshold}.

\myfigure{Threshold}{Different geometries (row) and the respective cast shadows under different direct shadow casting thresholds (columns).
The camera is to the left and the light in the back.
The top two geometries are common cylinders, but there is no single threshold (pink) that would result in the correct value for both cases.
Our method learns to make the decision adapted to the color and depth context.
The last row shows that the solution is ambiguous, as the depth map observed from the camera is identical to the wide cylinder.
Our network would likely learn to cast a shadow similar to a cylinder.}

While the task is ill-posed (\refFig{Threshold} last row), NNs have demonstrated a great ability at building deep priors to infer 3D geometry from only a single view-point \cite{henzler2019escaping,nguyen2019hologan}, given a large enough set of examples.
We often use such priors ourselves, for instance, a vertically elongated object like a lamp post might be an ``accidental'' view of a very long wall from the side (\refFig{Threshold} last row), but in expectation over a dataset, it is just what it is, a lamp post, casting a lamp post shadow.
Similarly, when observing the front of a building, we assume a thick filled volume behind it; we could be observing a movie set with fake buildings, but this is very unlikely.

\mypara{Possible solutions}
The most straightforward way to take into account the long-range interactions caused by occluding geometry given a depth map would thus be to ray-march the depth buffer to compute shadows \cite{ritschel2009approximating}.
More precisely, to shade a point, all pixels that fall on the ray toward the light source are considered.
If the depth buffer is sufficiently close (\eg under a specified threshold) to the corresponding depth of the 3D ray for any of these pixels, the point is considered in shadow.
Using a small threshold works well as long as the light direction is relatively orthogonal to the camera plane as most of the useful information is contained in the depth map.
However, these methods cannot properly deal with unobserved thickness when the light is coming from the side or the top of the image as, there, the threshold plays an important role (\refFig{Threshold}).
A real example illustrates this issue in \refFig{DirectProblem}.
The top of the aqueduct not being observed leads to underestimation of shadow regions.
We later refer to this approach as \method{Direct}.

On the other hand, if we were to accept shadow prediction to be a task worth learning, the most straightforward way would be to employ an image-to-image translation network from the depth and color image to the shadow image \cite{nalbach2017deepshading}.
We will indeed study this ablation (\method{2D}), but it has limited quality due to two main reasons.
Firstly, convolutions in an encoder-decoder take into account the 2D context at different scales, but for a shadowing task, the context that matters from a query point, in a certain direction, is 1D, along a ray.
Meaning that the relevant information can be arbitrarily far from the shadowed point.
Secondly, it does not make use of the physics of shadows.
They are a combination of attached shadows (also known as self shadows), which depends on local surface orientation and cast shadows, that are produced by the presence of an occluder that can be arbitrarily distant.

Accounting for the duality of shadows (self and detached shadows), the partial effectiveness of ray-marching and recent advances in deep learning inspired our solution.

\myfigure{DirectProblem}{Illustration of the problem caused by casting shadows using ray-marching directly in the depth map.
Left: an input image.
Right: a schematic representation of surface-based shadow casting.
While the plain brown zone is correctly classified as shadow, casting rays from the hatched one toward the sun does not produce intersections with the observed surface resulting in the incorrect classification of this zone as non-shadow.}

\mypara{Our solution}
Our solution computes both the attached and the detached shadow in a single network.
Instead of predicting binary shadow images, we use the product of the binary mask with the cosine term (that we later refer to) as the new shadow image $\new\shadowImage$.

As shown in \refFig{ShadowCosine}, this has three main advantages.
Firstly, it prevents creating arbitrary high frequency cut-offs on smooth surfaces as the cosine term smoothly goes to zero before the mask does.
Secondly, it contains part of the surface shading, giving cues about direct light intensity change.
Finally, as we have two different lighting conditions, we also have information regarding the difference of intensity between shadow and non-shadow regions.
When the light is at a grazing angle (bottom row \refFig{ShadowCosine}), our representation encodes that the difference between shadow and non-shadow region is smaller than when the light is at a higher angle (top row \refFig{ShadowCosine}).

\myfigure{ShadowCosine}{From left to right: A rendering of an example scene, binary shadow images and our proposed representation: cosine terms times shadow images. We show two illumination directions, a high lighting (top) and lighting at a grazing angle (bottom).}

To predict a shadow image we first observe that we can provide to the network an approximation of the cosine term, thus helping with attached shadows, by computing the clamped dot product of the light direction and the approximate normal image $\max(0, \left<N, \new\lightDirection\right>)$.
This guide is therefore as inaccurate as the normal map and is a prior that will need to be refined by the network.
To do so we concatenate this term with the original RGB image.
We later refer to this four channels image as the ``2D features'':

\begin{equation}
2D_\mathrm{feat}= \mathtt{cat}[\old\colorImage,\max(0, \left<\normalImage, \new\lightDirection\right>)]
\end{equation}

As discussed, shadows \emph{cast} from arbitrarily distant occluders are more difficult to handle. The key insight of this paper is that the information to be considered for deciding if a point is in cast shadow relies on all the image positions, and their respective local neighbors, that fall onto a ray from that point in the direction to the light, \ie the same pixels used for casting shadows with ray-marching.

\begin{figure}[h!]
    \centering
    \includegraphics[width=0.4\linewidth]{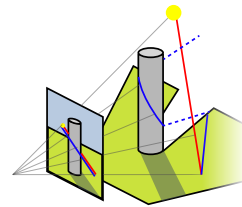}
    \vspace{-.2cm}
    \caption{A cylinder casting a shadow and its projection. The red line shows the depth of the ray while the blue line shows the depth sample in 2D ray-marching.}
    \label{fig:Solution}
\end{figure}

\refFig{Solution} shows a cylinder and two associated depths along a light ray, the one from the observed surface (blue) and the one from the 3D ray (red).
To answer if a point is in shadow, we have to consider all points on a ray from that point to the light.
The most relevant information available to evaluate this is in the depth and color image along the geometry of that ray.
An ideal method would classify points as occluded or not occluded, but using depth information directly, results in the errors previously mentioned seen in \refFig{DirectProblem} if ray-marching against the implied surface.
Instead, we also rely on color, as well as on nearby depth values to account for the full spatial arrangement. 
Below we detail this process.

\mycfigure{Marching}{Our ray-marching procedure.
Left (a), we show 4 pixels marched in the RGB and depth map into the direction of the light with 8 steps.
For that row, this results in a 3D image of 8 layers, one for each ray-marching step.
This epipolar volume is fed into a 3D encoder branch that reduces all dimensions by half in three steps while doubling feature count (b).
We opt for this approach to allow the network to take into account immediate local neighbors near the ray but restrict the network from computing spatially global features across the image plane.
This is followed by steps that keep spatial resolution in the volume height and width, but reduce the depth dimension to 1.
This is decoded into a 2D image.
An additional 2D branch is provided the approximate Lambertian term, that is 2D-encoded.
On the decoding step the 2D and 3D branches share features
Orange lines indicate skip connections.}

For every image pixel, we ray-march $z=256$ steps in the 2D direction, defined by the projection of the 3D light direction.
We sample the input pixel depth and color values at that position and store them into separate channels.
Color values are stored as they have been shown to aid learning a per-pixel object thickness value which is important for shadow estimation \cite{nicastro2019x}.
Instead of storing depth directly, we store the ratio between the depth in the depth map (\refFig{Solution}, blue line) and the depth of the point along the ray (\refFig{Solution} red line).
This ratio is attractive as it is the quantity that a normal ray-marching would use to decide occlusion.
When this ratio is very close to 1, it means that the ray intersects the surface, when it is greater than one it means the ray is in front of the observed surface and behind it when the ratio is smaller than one, which is the case for the points on the cylinder in \refFig{Solution}.
Second, this ratio is scale-invariant. 
Scale-invariance is very important in our scenario as depth-estimators learn disparity, and not absolute depth values, so are themselves also scale independent. 

We stack all $z$ features, resulting from the $z$ steps of the ray-marching, into a volume of size $x\times y\times z\times4$, the four channels being RGB and depth ratios.
We refer to this volume as the 3D features $3D_{feat}$.
Similar ideas have been used to find correspondences between pairs of images, where for each point in one image, the epipolar line in the other image is marched \cite{he2020epipolar}.
Instead, here we march the epipolar image line of a point with respect to the light.

Finally, a new 3D-2D encoder-decoder with two encoder heads maps the $x\times y\times z\times4$ $3D_\mathrm{feat}$ volume to a bottleneck size of $x'\times y'\times 1\times n$ using 3D convolution in the 3D head.
Intuitively, applying 3D convolution to the proposed volume allows to traverse along the ray direction while also accounting for local context.

In parallel, the 2D features $2D_\mathrm{feat}$ are also encoded to a $x'\times y'\times n$ image using standard 2D convolutions.

Finally, a decoder maps the max of the encoded 2D and 3D features back to an image of size $x\times y$ with a single channel using 2D convolutions.
Our network architecture is illustrated in \refFig{Marching} and outlined in detail in the supplementary material.
For the skip connections from the 3D branch to the 2D decoder we use a linear upsampling layer to match the volume sizes.
This network is asked to match the ground truth new shadow image:
\begin{equation}
\loss_\cshadowNet(\old\colorImage,\new\lightDirection)=
\imageMetric(
\cshadowNet^\tunableParameters(
\old\colorImage,
\new\lightDirection)
,
\new\shadowImage
)
\end{equation}
as the loss function \imageMetric we use E-LPIPS \cite{kettunen2019lpips}, it led to more plausible outputs as it is much more resilient to small errors in shadow boundaries than \ac{MSE} which led to smooth shadow when we tested it.

\mysubsection{Extracting the original shadow}{Deshadow}
In order to get the source image shadow in the source light direction $\old\lightDirection$, we could simply execute $\cshadowNet$ for that direction \ie $\sshadowNet=\cshadowNet$.
Unlike the target direction $\new\lightDirection$ that is user defined, the original light direction $\old\lightDirection$ is unknown.
To obtain it at test time, we rely on a simple user interface which shows the output of $\cshadowNet$.
The user is asked to roughly align the predicted shadows with the ones in the input image.
We show a demonstration of this process in the supplemental video.
The interface could also be initialized with outdoor light estimation methods such as \cite{holdgeoffroy-cvpr-19,legendre2019deeplight}.
Furthermore, during training we artificially add noise to the ground truth light direction to make the model more robust to inaccuracies from the user.
Whilst running the input and guide images through the $\cshadowNet$ gives us a shadow image, it misses the opportunity to refine the shadow with the assumption that the very shadow we look for are present in the color image $\old\colorImage$, albeit entangled with albedo.
In light of this, to make use of both the shadow in the image, and the shadow predicted by the network, a separate network $\refineNetwork$ is trained, with the sole purpose of refining the output of $\cshadowNet$, while also accessing the color image, such that $
\sshadowNet(\old\colorImage,\old\lightDirection)=
\refineNetwork(\cshadowNet(\old\colorImage,\old\lightDirection),\old\colorImage,\old\lightDirection)$.
$\refineNetwork$ is supervised, with RGB \ac{MSE} as here the shadow boundaries are available in the original image and this image needs to be accurate for better shadow removal rather than just plausible:
\begin{equation}
\loss_\sshadowNet(\old\colorImage,\old\lightDirection)=
\imageMetric(
\sshadowNet^\tunableParameters(
\old\colorImage,
\old\lightDirection
)
,
\old\shadowImage
).
\end{equation}%
\mysubsection{Relighting}{Relighting}
With both shadow images at our disposal, we can now perform a relighting closely inspired by 
Philip et al., \cite{philip2019multi} which is a rich image(s)-to-image translation.

Input to the relighting network $\relightingNet$ are the two shadow images $\new\shadowImage$ $\old\shadowImage$, the old color image $\old\colorImage$, the computed normals $\normalImage$, as well as the old and new light direction $\old\lightDirection$ and $\new\lightDirection$ along a direction map $\Psi$ containing the $x,y,z$ direction from the camera center towards each pixel center.
The depth is not an input to the relighting network.
Output is the final color image $\new\colorImage$.
\refFig{Architecture} shows the complete network architecture.

The loss function penalizes the color image in the new light condition $\relightingNet(\old\colorImage,\old\lightDirection\new\lightDirection)$ to match a known reference color image $\new\colorImage$ in the new light conditions, so
\begin{equation}
\loss_\relightingNet(\old\colorImage,\old\lightDirection,\new\lightDirection)
=
\imageMetric(
\relightingNet^\tunableParameters(
\old\colorImage,
\old\lightDirection,
\new\lightDirection)
,
\new\colorImage
)
\end{equation}
where \imageMetric is E-LPIPS \cite{kettunen2019lpips}.

\mysubsection{Generating realistic images and shadows}{Adverserial}
Additionally, we employ a Patch-based \ac{LSGAN} \cite{mao2017least} with a 64 square pixel receptive field to enforce the results to align with the distribution of natural images.
Instead of conditioning it only on the output image, we also condition it on the original RGB image $\old\colorImage$ and the new shadow image $\new\shadowImage$.
This adversarial loss has three main effects.
Firstly, it helps with the overall visual quality and sharpness of the images.
Secondly, it improves shadow removal.
Our intuition is that by seeing the input and output image the discriminator should be able to detect bad shadow removal.
Lastly, having the shadow image as part of the conditioning helps with shadow details and coherency between predicted shadows and the output image.

\mycfigure{Buffers}{Visualization of different intermediate buffers for a real image relighting example. Top row, from left to right: Input image $\old\colorImage$, Estimated Depth $\depthImage$, Normals $\normalImage$.
Bottom row, from left to right: Estimated Source Shadows $\cshadowNet(\old\colorImage,\old\lightDirection)$, 
Refined Source Shadows $\sshadowNet(\old\colorImage,\old\lightDirection)$, 
Estimated Target Shadows $\cshadowNet(\old\colorImage,\new\lightDirection)$, 
Output $\new\colorImage$.}

\mysubsection{Training data}{TrainingData}
Creating a real-world training dataset for our method with the full distribution of lighting conditions our model accounts for would be exceptionally challenging. 
Instead, we train our method entirely on synthetic scenes.
This enables us to simulate the full spectrum of lighting conditions in a large variety of settings.
We utilize 40 Evermotion Archexterior \cite{evermotion} scenes for geometry, material and texture of the scenes.

For each scene a camera path is manually created, from which we sample 256 view-points.
For each view-point we select an object for the camera to look at, a random focal length and render the scene under 16 lighting conditions at 1024$\times$768 resolution.
In total we render 164k images from the 40 scenes at 128 samples per pixel using the Cycles \cite{blender} path tracer.
For lighting, all images are rendered using the Nishita sky model \cite{nishita1993display} which implements atmospheric scattering to which we added volumetric clouds to handle their relighting in real images.
Using this realistic sky model, clouds and the direction map $\Psi$ allows the relighting network to implicitly detect and relight the sky.
$\Psi$ is particularly helpful in removing and synthesizing the sun when it is directly visible in the image as the network can match the old or new sun direction $\old\lightDirection$ and $\new\lightDirection$ with the direction towards which each pixel points, which is given by $\Psi$.
An illustration of such synthesis and cloud handling is visible in \refFig{teaser}, ``Relighting A''.

On top of the path-traced RGB image, each individual sample contains a ground truth depth map, normal image and its corresponding shadow image. This shadow image is rendered by computing single bounce direct illumination of the scene, replacing all the materials with white Lambertian BRDF, as shown in \refFig{ShadowCosine}, third column.
Training examples are are also shown in the supplemental materials.

\mypara{Out of screen shadow casters} Our ray-marching based model is not able to handle shadows cast by objects that are not visible in the image.
Thus, we cull all the geometry outside of the screen before rendering each training image.
This means none of our training examples exhibits shadows cast by out of screen objects.
While at test time the original image may contain such shadows, we found that the network has learnt to remove them correctly. 
As for the new shadows, we empirically find that synthesizing shadows cast only by the visible content provides sufficient realism in most cases.

\mysubsection{Training strategy and details}{TrainingStrategy}
\mypara{Training procedure}
At train time we sample a viewpoint from the dataset of rendered scenes and a random pair of lighting conditions to define the input $\old\colorImage$ and target image $\new\colorImage$.
These images are stored in linear space and similar random exposure, saturation and gamma tone-mapping augmentations are applied to both $\old\colorImage$ and $\new\colorImage$.

\mypara{Learning to trust the depth}
As previously mentioned, predicted depth maps are often distorted.
This means that shadows cast by predicted depth maps are often miss-aligned with the ground truth ones, though they may look realistic.
When training our cast shadow network only with predicted depth maps, this phenomenon led to very noisy gradients and poor convergence quality.
Trying to correct this distortion would be equivalent to trying to beat the best monocular depth estimator.
Instead, we teach the cast shadow network to trust the depth maps by training it with a combination of ground truth and NN estimated depth maps.
Each training step flips a 80\,\%-to-20\,\%-biased coin to determine if it learns new cast shadows from ground truth or estimated depth respectively.
Both the old shadows and normals are always computed on the estimated depth maps.
In doing this, our cast shadow network learns geometrically founded features (\eg the relationship between surface normals, light direction and attached shadows), however, is also robust to noisy normals and inaccurate cast shadows computed for the old light direction $\old\lightDirection$ and allows to transfer to real data more easily.

\mycfigure{MainResult}{Results of our proposed approach for three novel illuminations (three right columns), on a variety of challenging real-word scenes (first column).
For more results, including full time lapse videos, please see the supplemental video.}

\mypara{Optimization details}
For the optimization we use the Adam Optimizer \cite{Kingma2014_adam_solver} with a learning rate of $1e-4$ both for the generator and the discriminator.
We alternate five steps of generator for one step of discriminator.
The loss is weighted so each sub-network has approximately equivalent losses at the start of training, $\lossWeight_\cshadowNet=10$,$\lossWeight_\sshadowNet=2$,$\lossWeight_\relightingNet=10$, the adversarial loss is set lower with $\lossWeight_\advNet=0.1$.
We train on 384x384px images and a batch size of 4 patches.

An important detail regarding the optimization procedure is visible in \refFig{Architecture}.
The small scissors denote that the gradients do not flow backward from the refinement network $\sshadowNet$ to the cast shadow network $\cshadowNet$.
If we were not to do this, half of the gradients of $\cshadowNet$ are from predicting the original shadows and $\cshadowNet$ learns to copy the shadows from the original image $\old\colorImage$.

\mypara{Network Architectures}
The 3D branch of the cast shadow network \cshadowNet is composed of 3D convolutions with kernel sizes (4,4,4) allowing to effectively divide by 8 the volume size after each convolution while we multiply the number of features by 2. After 3 down-samplings, additional convolutions with kernel size (4,3,3) reduce the depth dimension to 1.

All other encoder and decoders are \ac{CNN} inspired by the architecture from Pix2Pix HD \cite{wang2018pix2pixHD}.
They are U-Net-like \cite{ronneberger2015u}, but composed of residual blocks and residual skip connections.

A detailed blueprint of the system with network architectures is available in the supplemental materials.

\mysection{Experiments and Results}{Results}

We test our approach on both a set of reserved synthetic scenes, allowing for quantitative assessment on ground truth data (\refSec{Quantitative}), as well as qualitatively on real-world images (\refSec{Qualitative}).
Our experiments are formed on a set of ablations of our method, justifying the benefit of our proposed learned 3D ray-marching module.

\mysubsection{Qualitative}{Qualitative}
We tested our method on a range of photos, typically of outdoor architecture shown in \refFig{MainResult}.

\mycfigure{Comparison}{Comparison of our methods against Self-supervised Outdoor Relighting \cite{yuye2020} and \cite{philip2019multi}.
\method{Ours} performs favorably over \method{Self Relighting} in all scenarios.
We perform on-par to Philip et al., \cite{philip2019multi}, however, do not require multi-view inputs.
Self-relighting is unable to remove the old shadow, as seen particularly in the top scene.}

\refFig{Comparison} shows examples of results for all methods we compare to, giving qualitative insight into the performance of our approach.
We perform favorably to \method{Self Relighting} in all scenarios.
Most notably is our ability to predict detached (cast) shadows and our ability to remove source shadows.
When comparing to Philip et al.,\cite{philip2019multi} we find our method to produce stronger shadows in the target image.

In \refFig{Art}, we apply our method to paintings.
Despite the shadows not always been physically accurate, our network still creates plausible outputs.
The ability for our network to perform well in such a large domain of inputs highlights its generalization abilities.

\mycfigure{Art}{Application to artwork.
The first column shows the original, the two left columns our result.
Both inputs show strong, while not yet entirely physically-correct shadow.
The top one is ``Odysseus returns Chryseis to Her Father'' (ca. 1644) by Claude Lorraine (1604--1682).
The bottom one is ``The square of Saint Mark's, Venice'' (ca. 1723) by Giovanni Antonio Canal (1697--1768) known as Canaletto.
}

\mysubsection{Quantitative}{Quantitative}

In this section we perform an ablation study on our method.
All test samples are drawn from scenes that were not available at training time.
This eliminates the chance the network will have seen similar view-points of the test images at train time.
We describe our ablation methods below.

\mypara{\method{Pix2PixHD-like}} An image-to-image translation network.
The source image $\old\colorImage$ is directly mapped to the target image $\new\colorImage$ using a \ac{CNN} U-Net architecture \cite{isola2017pix2pix}.
In practice, for fairness, we use our relighting network $\relightingNet$ with all its inputs except for the computed old and new shadow images.
The relighting approach is defined as $\new\colorImage = \bar\relightingNet(\old\colorImage, \old\lightDirection, \new\lightDirection)$.

\mypara{\method{2D}} The above method ignores any shadow guides for training.
Here, we replace our 3D shadow estimation network $\cshadowNet$ with a 2D \ac{CNN} similar to the refinement network $\sshadowNet$, its input are the input image $\old\colorImage$, light direction $\old\lightDirection$ or $\new\lightDirection$, the corresponding cosine term and the estimated depth map.

\mypara{\method{Direct}} We replace our 3D shadow network with a non-learnable direct screen-space ray marching algorithm \cite{ritschel2009approximating}.
Shadow images are computed directly from the estimated depth map and input to the final relighting network.
The samples collected for this method are the same samples passed into our learnable 3D shadow network.

\mypara{\method{Our}} Our full method, as described in \refSec{OurApproach}.

All methods make use of depth extracted by $\depthNet$, as described in \refSec{OurApproach} and no method has access to the ground truth depth or normals, only to RGB and source light direction $\old\lightDirection$.

\mypara{Metrics}
As test data is rendered, we know the correct new image $\new\colorImage'$ and hence can compute the image error, here using Structural Dissimilarity Index (DSSIM), Mean Squared Error (MSE) and Perceptual Similarity (LPIPS).
To further evaluate the networks ability to predict shadows, we also report the MSE of the predicted and ground truth shadow images.
Note, we cannot do this for \method{Pix2Pix} as no intermediate shadows are produced.

\mycfigure{Ablation}{Qualitative results of our ablation study. In general \method{Pix2PixHD-like} achieves reasonable global relighting, however, performs poorly at both shadow removal and placement. 
\method{2D} is not capable of long-range detach shadows as standard 2D convolutions are unsuitable for such a task. 
\method{Direct} is the most similar to \method{Our} method, however, suffers from underestimated shadows.}

\mypara{Results}
Results are summarized in \refTbl{Results}.
We see that our approach performs best according to all metrics, both in the image error, as well as for the intermediate shadow.
Additional visual evidence is shown in \refFig{Ablation} and discussed in the caption.

\begin{table}[]
    \small
    \centering
    \caption{Relighting error for different ablations across the test dataset according to different metrics.
    For all methods with shadow prediction, we also report the MSE.
    For all metrics, less is better.}
    \label{tbl:Results}
    \renewcommand{\tabcolsep}{3pt}
    \begin{tabular}{lrrrr}
        \toprule
        \multicolumn1c{Method}&
        \multicolumn3c{Relight}&
        Shadow\\
        \cmidrule(lr){2-4}
        & 
        \multicolumn1c{DSSIM$\downarrow$} & 
        \multicolumn1c{LPIPS$\downarrow$}& 
        \multicolumn1c{MSE$\downarrow$}&
        \multicolumn1c{MSE$\downarrow$}\\
        \midrule
        \method{Pix2Pix}&  .165& .0170& .0427& \multicolumn1c{---} \\
        \method{2D}&  .169& .0182& .0515& .187 \\
        \method{Direct}& .157& .0169& .0409& .358 \\
        \method{Our}&  \textbf{.154}& \textbf{.0160}& \textbf{.0399}& \textbf{.175} \\
        \bottomrule
    \end{tabular}
\end{table}

\mysection{Limitations}{Limitations}

Our method is subject to several assumptions.
Firstly, we require the input image light direction.
While we use a manually derived input for the results shown in this paper, the proposed system would benefit from more reliable methods to estimate this.

In all scenarios, we assume a dominant single light source, typically, the sun.
Extension to a mixture of light sources is straight-forward (as light sums linearly) but a more refined solution, \eg a latent model of illumination, will certainly outperform this.

Currently, we model classic opaque shadows in our ray-marching.
Hence, all other shading effects, like reflections, colored shadows, indirect light or caustics are not modeled explicitly, but instead left to the final shading network to approximate.
This works, as long as such effects do not become visually dominant.
Relighting a modern office interior (mirror reflections), a church interior (colored shadows), a glass vase (caustics) or strong indirect lighting would require both adequate training data, but also likely require adapted guide signals, like we provided for shadows.
Whilst our method does not allow for control of shadow softness, this would be an easy extension, as shadow softness can be controlled in the training data.

As discussed in \refSec{TrainingData}, our method assumes there are no shadows cast from out of screen geometry. 
Although it is conceivable that such geometries can be reconstructed from their shadows alone, we do not address such scenarios in this work.

Finally, whilst all the results presented in this work are generated using a single model state (including \refFig{Art}), our network components are learnt through a data-driven approach. 
Therefore, for some test images where the source image is far from the training distribution our results degenerate. 
This is, however, more a limitation of our training data than the system architecture.

\mysection{Conclusion}{Conclusion}
We have shown how the classic idea of ray marching, combined with a learned component, allows to cast shadows in RGB images resulting in faithful outdoor relighting of single-view images.
This is made possible by using geometry provided from an off-the-shelf monocular object detector.
Our method compares favorably to previous work as well as to strong baselines of ablations. 
When looking at the evolution of screen space shading however, it appears quite conceivable that the idea of a (differently) ray-marched guide is applicable to all of these in future work.
Beyond relighting, handling physically based long-range interactions from a single image, our key technical innovation, might have applications in other graphics tasks such as image material editing and even lead to novel forms of (self) supervision using physical long-range constraints in both vision and graphics.

\bibliographystyle{eg-alpha}
\bibliography{paper}

\clearpage
\input{appendix}

\end{document}

%% file: commands.tex
\usepackage{graphicx}
\usepackage{soul}
\usepackage{amsmath}
\usepackage{amssymb}
\usepackage{microtype}
\usepackage{wrapfig}
\usepackage{xcolor}
\usepackage{booktabs}
\usepackage{multirow}
\usepackage{xspace}

\def\figurePath{Figures/}

\def\myfigure#1#2{%
    \begin{figure}[htb]%
    \centering\includegraphics*[width = \linewidth]{\figurePath#1}%
    \vspace{-.2cm}%
    \caption{#2}%
    \label{fig:#1}%
    \end{figure}%
}

\def\mycfigure#1#2{%
    \begin{figure*}[h!]%
    \centering\includegraphics*[width = \linewidth]{\figurePath#1}%
    \vspace{-.2cm}%
    \caption{#2}%
    \label{fig:#1}%
    \end{figure*}%
}

\newcommand{\eg}{e.\,g.,\ }
\newcommand{\ie}{i.\,e.,\ }
\newcommand{\etal}{et~al.\ }

\newcommand{\refSec}[1]{Sec.~\ref{sec:#1}}
\newcommand{\refFig}[1]{Fig.~\ref{fig:#1}}

\newcommand{\refTbl}[1]{Tbl.~\ref{tbl:#1}}
\newcommand{\refAlg}[1]{Alg.~\ref{alg:#1}}

\soulregister\ref7
\soulregister\cite7
\soulregister\refFig7
\soulregister\refAlg7
\soulregister\cite7
\soulregister\ref7
\soulregister\pageref7
\soulregister\shortcite7
\soulregister\eg0
\soulregister\ie0
\soulregister\etal0

\DeclareGraphicsExtensions{.png,.jpg,.pdf,.ai,.psd}
\newcommand{\mysection}[2]{\section{#1}\label{sec:#2}}
\newcommand{\mysubsection}[2]{\subsection{#1}\label{sec:#2}}

\newcommand{\mymath}[2]{\newcommand{#1}{\TextOrMath{$#2$\xspace}{#2}}}

%% file: appendix.tex
\mysection{Supplementary Material}{Supplementary}

\mysubsection{Detailed Network Architecture}{Arch}
In \refFig{DetailedArch} we present a detailed architecture blueprint of our network with all inputs, modules, networks, and outputs.

\mycfigure{DetailedArch}{Blueprint of our full relighting system. Where arrows are not included assume flow remains in the current direction. All deep learning-based functions were implemented in the PyTorch framework (v1.9).}

\mysubsection{Training Data Examples}{DataEx}

In \refFig{Data}, we present random training examples from our dataset. For each viewpoint we provide two lighting conditions and their respective ground truth shadows.

\mycfigure{Data}{Examples of training pairs. From left to right: Input, Ground Truth Source Shadows, Target, Ground Truth Target Shadows.}

\mysubsection{Real world ground truth evaluation}{GroundTruth}
We evaluate our method on a real world lighting scenario. To enable this we utilize separate images taken roughly from the same viewpoint with different lighting conditions.
As shown in \refFig{GTCompare} we observe that whilst the network output looks plausible, the shadows are misaligned from the target ground truth.
We believe this is due to distortions in the predicted depth estimation for this scene.

\mycfigure{GTCompare}{Comparison to real images with different lighting conditions. From left to right: Input, Our relighting, Ground Truth.}

\begin{table}[]
    \small
    \centering
    \caption{Relighting error for different loss related ablations across the test dataset according to different metrics.
    We also report the MSE for shadow prediction.
    For all metrics, less is better.}
    \label{tbl:Ablations}
    \renewcommand{\tabcolsep}{3pt}
    \begin{tabular}{lrrrr}
        \toprule
        \multicolumn1c{Method}&
        \multicolumn3c{Relight}&
        Shadow\\
        \cmidrule(lr){2-4}
        & 
        \multicolumn1c{DSSIM$\downarrow$} & 
        \multicolumn1c{LPIPS$\downarrow$}& 
        \multicolumn1c{MSE$\downarrow$}&
        \multicolumn1c{MSE$\downarrow$}\\
        \midrule
        MSE loss for target\\ shadows (no LPIPS)&  .171& \textbf{.0159}& .0437& \textbf{.150}\\
        No PatchGAN& .162& .0190& .0405& .169\\
        Our& \textbf{.154}& .0160& \textbf{.0399}& .175\\
        \bottomrule
    \end{tabular}
\end{table}

\mysubsection{Further Ablations}{FurtherAblation}

In addition to the ablations presented in the main paper (Sec. 5.1 and Sec. 5.2), we also undertake further ablations specifically evaluating the use of specific loss function components.
\refTbl{Ablations} provides the quantitative evaluation for these ablation.
As expected, the MSE loss is smaller for shadows when used as the training metric. Overall the pipeline appears to perform marginally better when the PatchGAN loss term is used which seems surprising. While small variations might be due to the randomness of the training process, it is possible that the PatchGAN loss helps escape local minima leading to better convergence.
Moreover, as shown in \refFig{GANnoGAN}, this loss helps in producing complex localized effects such as high frequency reflections on tree leaves (\refFig{GANnoGAN} first row) or reflection on the water and better looking clouds (\refFig{GANnoGAN} second row).
Training with E-LPIPS \cite{kettunen2019lpips} for shadows does not provide a strong advantage numerically but has a strong impact on the sharpness on elongated shadows as can be seen in \refFig{LPIPS}. Without the E-LPIPS loss and with a more traditional MSE loss, the shadow network tends to produce overly smooth shadows as a slight misalignment of boundaries is strongly penalized.

\mycfigure{GANnoGAN}{Ablation results when removing the PatchGAN loss. From left to right: Input, Our relighting, Ablation result.}

\mycfigure{LPIPS}{Ablation results where the E-LPIPS loss is replaced by an MSE loss. Input to the network is shown on the left. (\textbf{top}) Ours for target shadow (left) and output (right). (\textbf{bottom}) Ablation for target shadow (left) and output (right).}